\documentclass[a4paper]{article}%
\usepackage{amsmath}
\usepackage{amsthm}
\usepackage{amsfonts}
\usepackage{amssymb}
\usepackage{graphicx}
\usepackage{enumerate}
\usepackage{rotating}
\usepackage{float}
\usepackage{pbox}
\usepackage{epstopdf}
\usepackage{multirow}
\usepackage[table]{xcolor}
\usepackage{threeparttable}
\usepackage{booktabs}
\usepackage{caption}
\usepackage{subcaption}
\usepackage[left=25mm,right=25mm,top=25mm,bottom=25mm]{geometry}%
\providecommand{\U}[1]{\protect\rule{.1in}{.1in}}
\providecommand{\U}[1]{\protect\rule{.1in}{.1in}}
\newtheorem{theorem}{Theorem}

\newtheorem{claim}[theorem]{Claim}
\providecommand{\U}[1]{\protect\rule{.1in}{.1in}}

\begin{document}

\title{Reproducing the Kolmogorov spectrum of turbulence with a hierarchical linear cascade model}
\author{Tam\'{a}s Kalm\'{a}r-Nagy\thanks{prl@kalmarnagy.com},\hspace{1mm} Bendeg\'{u}z
Dezs\H{o} Bak\thanks{bak@ara.bme.hu}\\Department of Fluid Mechanics, Faculty of Mechanical Engineering, \\Budapest University of Technology and Economics}
\date{}
\maketitle

\begin{abstract}
In Richardson's cascade description of turbulence, large vortices break up to
form smaller ones, transferring the kinetic energy towards smaller scales.
Energy dissipation occurs at the smallest scales due to viscosity. We study
this energy cascade in a phenomenological model of vortex breakdown. The model
is a binary tree of decreasing masses connected by softening springs, with
dampers acting on the lowest level. The masses and stiffnesses between levels
change according to a power law. The different levels represent different
scales, enabling the definition of \textquotedblleft mass
wavenumbers\textquotedblright. The eigenvalue distribution of the model
exhibits a devil's staircase self-similarity. The energy spectrum of the model
(defined as the energy distribution among the different mass wavenumber) is
derived in the asymptotic limit. A decimation procedure is applied to replace
the model with an equivalent chain oscillator. For a range of stiffness
parameter the energy spectrum is qualitatively similar to the Kolmogorov
spectrum of 3D homogeneous, isotropic turbulence and we find the stiffness
parameter for which the energy spectrum has the well-known $-5/3$ scaling exponent.

\end{abstract}

\section{Introduction}

Many natural phenomena and engineering processes exhibit energy transfer among
a range of different scales. Frequently studied examples include nonlinear
chain oscillators
\cite{fermi1955studies,gendelman2001energy,vakakis2001energy,
vakakis2008nonlinear}. Direct energy cascade describes primary energy transfer
from large scales to small ones, while inverse cascade
\cite{turcotte1999inverse} refers to energy transfer from small scales towards
larger ones. Turbulent flow is a prime example of a process which exhibits
different scales and an energy cascade. Richardson's so-called "eddy
hypothesis" \cite{richardson1920supply, richardson1922weather} argues that the
largest eddies are unstable and break up forming several smaller vortices,
gradually transferring the kinetic energy of the flow to smaller scales.

\begin{figure}[h]
\center
\includegraphics[width=8.6cm]{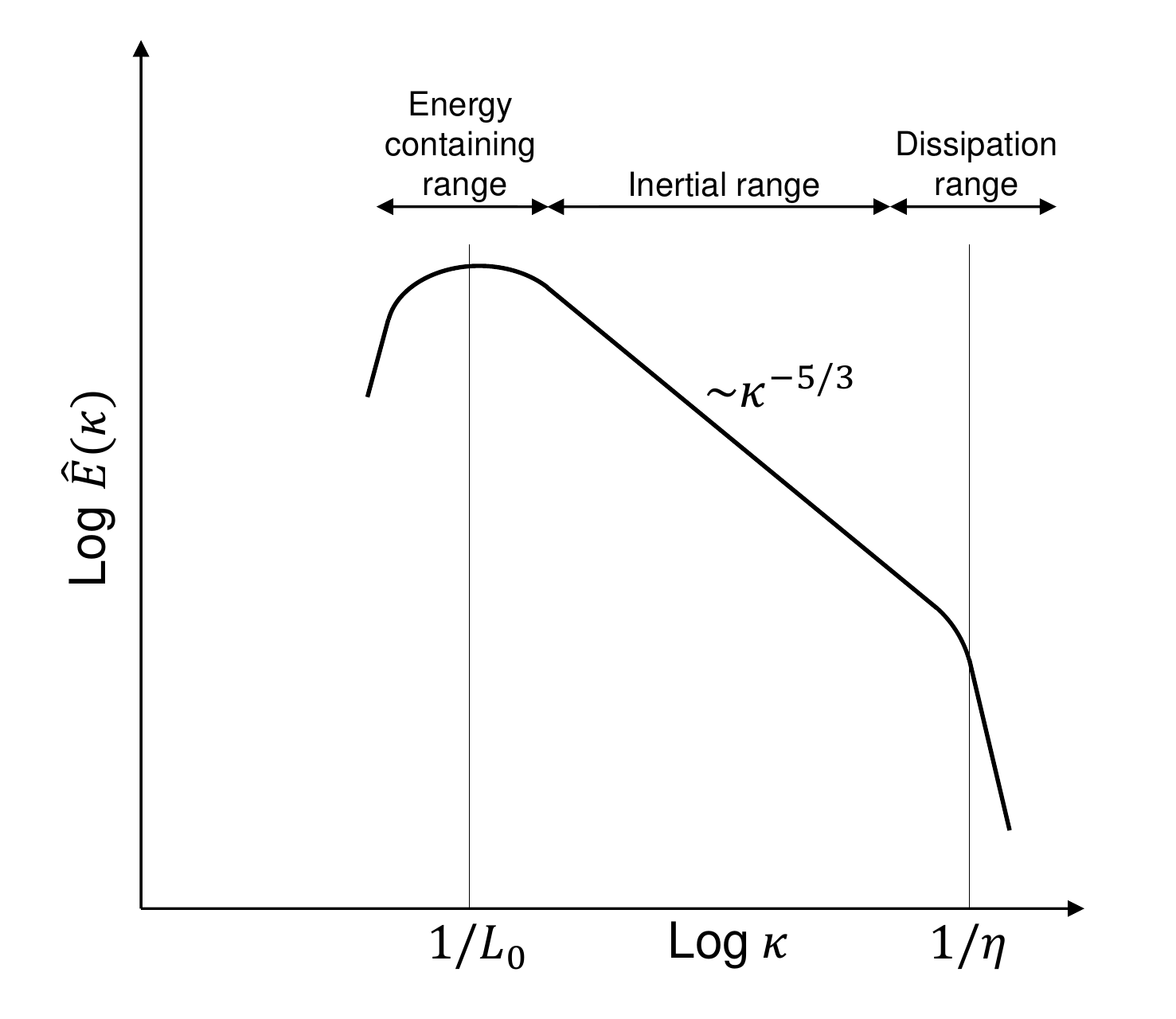}\caption{Qualitative graph of
the Kolmogorov spectrum.}%
\label{kolmogorovspectrum}%
\end{figure}

The turbulent energy cascade is characterized by the energy spectrum $\hat
{E}(\kappa)$ which describes the distribution of the total energy
\begin{equation}
E=\int\hat{E}(\kappa)d\kappa.\label{espectrum}%
\end{equation}
among the different scales. \noindent The wavenumber $\kappa\sim1/L$ is
associated with the vortex having characteristic size $L$. In Figure
\ref{kolmogorovspectrum} the so-called Kolmogorov spectrum of 3D homogeneous
isotropic turbulence is shown, illustrating the main features of the turbulent
energy cascade
\cite{kolmogorov1941local,pope2000turbulent,ditlevsen2010turbulence}. The
Kolmogorov length scale is denoted with $\eta$.

\noindent The energy production mainly affects the large scales
(characterized by $L_{0}$), hence the bulk of the energy is contained in the
large eddies (energy containing range). In the intermediate wavenumbers
($1/L_{0}<\kappa<1/\eta$, called the inertial range) the energy spectrum is
described by the famous Kolmogorov scaling law
\cite{kolmogorov1941local,ditlevsen2010turbulence}
\begin{equation}
\hat{E}(\kappa)\sim\kappa^{-5/3}.\label{kolmspectrum}%
\end{equation}
\noindent The largest wavenumbers ($\kappa>1/\eta$) are associated with length
scales smaller than the Kolmogorov scale. In this dissipation range most
kinetic energy is lost due to viscous friction. The Kolmogorov spectrum is
confirmed by many measurements and simulations \cite{ertuncc2010homogeneity,
kang2003decaying, galanti2004turbulence, biferale2003shell}.

\begin{figure}[h]
\centering
\begin{subfigure}[h]{4.3cm}
\includegraphics[width=4.3cm]{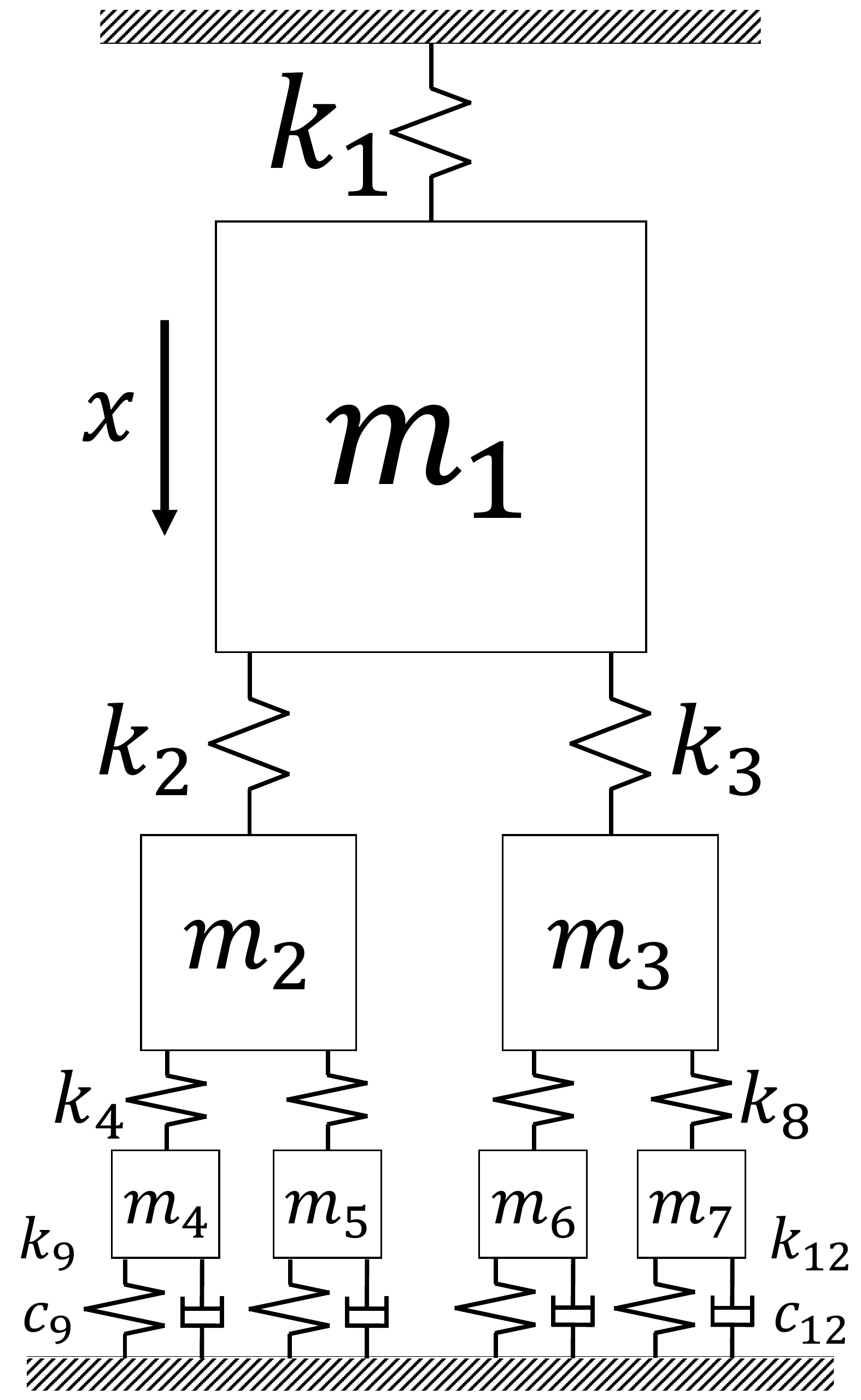}
\caption{}
\label{mechmodandchainosca}
\end{subfigure}
\begin{subfigure}[h]{4.3cm}
\includegraphics[width=4.3cm]{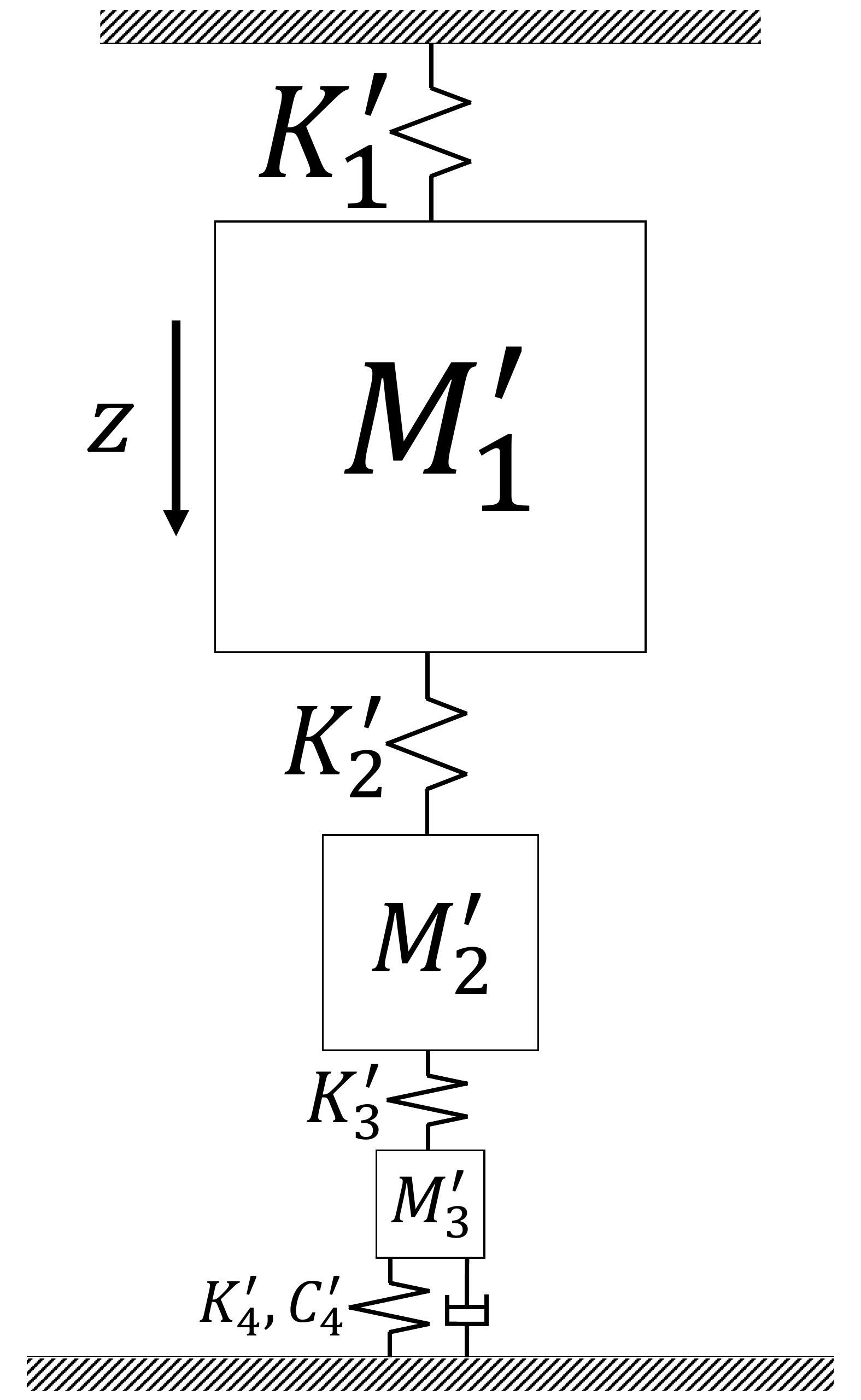}
\caption{}
\label{mechmodandchainoscb}
\end{subfigure}
\caption{(a) The mechanistic model and (b) the
equivalent chain oscillator.}%
\label{mechmodandchainosc}%
\end{figure}

We present a phenomenological model of turbulence inspired by Richardson's
cascade description and to demonstrate that this model exhibits the Kolmogorov
spectrum for a suitably chosen parameter. The model is a binary tree of masses
and springs (Figure \ref{mechmodandchainosca}) in which the masses represent
the different scales, and the springs are responsible for the energy transfer.
\noindent Even though this model is linear, linear systems can
also contain a wide range of scales and exhibit complex behavior
\cite{kalmar2017complexity}. In fact, nonlinear systems can be embedded in
infinite-dimensional linear systems \cite{kowalski1991nonlinear}. Leyden and
Goodwine \cite{leyden2018fractional} set up a similarly structured
mass-spring-damper system and showed that the infinite version of the system
is described by a fractional ($1/2$th order) differential equation. Many
studies aim at designing effective vibration absorbers by realizing targeted
energy transfer, a primarily one-way energy transfer from the primary system
to the dissipative element. Tripathi et al. \cite{tripathi2017optimal}
investigated and compared the performance of a nonlinear energy sink (a light
mass with viscous damping and an essentially nonlinear spring) and linear
tuned mass damper on multi-degree-of-freedom systems. Pumh\"{o}ssel found a
method to effectively transfer energy among vibration modes using impulsive
force excitations \cite{pumhossel2016suppressing}.

\section{Mechanistic turbulence model}

\label{mechmodel}

Our mechanistic turbulence model is an $n$-level binary tree of masses and
springs, with dampers at the ground level (Figure \ref{mechmodandchainosca}
shows a $3$-level model). Level $l\in\left\{  1,...,n\right\}  $ consists of
$2^{l-1}$ masses, the total number of masses is $N=2^{n}-1$. The top mass is
connected to the rigid ceiling with a spring, while the bottom masses are
connected to the rigid ground with springs and dampers (signifying dissipation
only at the lowest scale). Every mass $m_{i}$ ($i=1,...N$) has a parent and
two (left and right) children whose indices are ($\lfloor.\rfloor$ denotes the
floor operation)
\begin{equation}
\mathcal{P}(i)=\lfloor i/2\rfloor,\quad\mathcal{L}(i)=2i,\quad\mathcal{R}%
(i)=2i+1,\quad i=1,...N.\label{parentchildren}%
\end{equation}
\noindent The equations of motion are
\begin{equation}%
\begin{split}
m_{i}\ddot{x}_{i}=  &  k_{i}(x_{\mathcal{P}(i)}-x_{i})+k_{\mathcal{L}%
(i)}(x_{\mathcal{L}(i)}-x_{i})+k_{\mathcal{R}(i)}(x_{\mathcal{R}(i)}-x_{i})+\\
&  +c_{i}\left(  \dot{x}_{\mathcal{P}(i)}-\dot{x}_{i}\right)  +c_{\mathcal{L}%
(i)}(\dot{x}_{\mathcal{L}(i)}-\dot{x}_{i})+c_{\mathcal{R}(i)}(\dot
{x}_{\mathcal{R}(i)}-\dot{x}_{i}),\quad i=1,...,N,
\end{split}
\label{eq:base}%
\end{equation}
\noindent with the boundary conditions (the ceiling and the ground are
motionless)
\begin{equation}
x_{i}=0,\quad i=0,N+1,...,2N+1.
\end{equation}
The initial conditions are%
\begin{equation}
x_{i}(0)=x_{i,0},\ \dot{x}_{i}(0)=v_{i,0},\quad i=1,...,N.\label{ICs}%
\end{equation}
Eqs. \eqref{eq:base} and \eqref{ICs} in matrix form is
\begin{equation}
\mathcal{M}\ddot{\mathbf{x}}+\mathcal{C}\dot{\mathbf{x}}+\mathcal{K}%
\mathbf{x}=\mathbf{0},\quad\mathbf{x}(0)=\mathbf{x}_{0},\quad\mathbf{\dot{x}%
}(0)=\mathbf{v}_{0},\label{eq:baseMtx}%
\end{equation}
\noindent where $\mathbf{x}$ is the vector of displacements, $\mathcal{M}$,
$\mathcal{C}$, $\mathcal{K}$ are the mass, damping, and stiffness matrices, respectively.

\subsection{Model parameters}

We introduce a quantity analogous with the wavenumber of a turbulent scale,
the \textquotedblleft mass wavenumber\textquotedblright\
\begin{equation}
\kappa_{l}=1/M_{l},\quad l=1,...,n.\label{masswavenumber}%
\end{equation}
\noindent Here $M_{l}$ is the mass scale representing masses in level $l$
(average, for example). For the ease of exposition we set every mass,
stiffness and damping coefficient equal within a level $l$ and denote these
with $M_{l}$, $K_{l}$, $C_{l}$. To represent different scales, the masses are
gradually decreased in lower levels, analogously to the eddy hypothesis, in
which a large vortex breaks up into smaller ones.

The mass scale $M_{l}$ is specified by the power-law distribution%
\begin{equation}
M_{l}=\left(  \frac{1}{2}\right)  ^{l-1},\quad l=1,...,n,\label{powerlawm}%
\end{equation}
\noindent thus the sum of masses in each level is 1. Similarly, we denote by
$K_{l}$ the stiffness coefficient of the springs connecting the masses of the
$l-1$th and $l$th levels. The $K_{l}$ values are also specified with a
power-law distribution with base $\sigma>0$ (stiffness parameter), i.e.
\begin{equation}
K_{l}=\sigma^{l-1},\quad l=1,...,n+1.\label{powerlawk}%
\end{equation}
\noindent By design, we only have dampers between the bottom masses and the
rigid ground, hence only $C_{n+1}$ is nonzero and we set
\begin{equation}
C_{n+1}=2\sqrt{M_{n}K_{n+1}}=2^{\frac{3-n}{2}}\sigma^{\frac{n}{2}%
}.\label{dimdamplarge}%
\end{equation}
\noindent The characteristics of the energy spectrum of the model (defined in
Section \ref{espectrumsection}) of course depends on the relation between the
base parameters of Eqs. \eqref{powerlawm} and \eqref{powerlawk} ($1/2$ and
$\sigma$). This means that one such base parameter is enough to describe the
behavior of the mechanistic model. This is why the base of the mass
distribution was set to $1/2$. Furthermore, based on our experience the energy
spectrum is practically independent of the damping.

\section{The eigenvalue distribution of the mechanistic model}

\label{chareq}

\begin{figure}[h]
\centering
\begin{subfigure}{8.6cm}
\includegraphics[width=8.6cm]{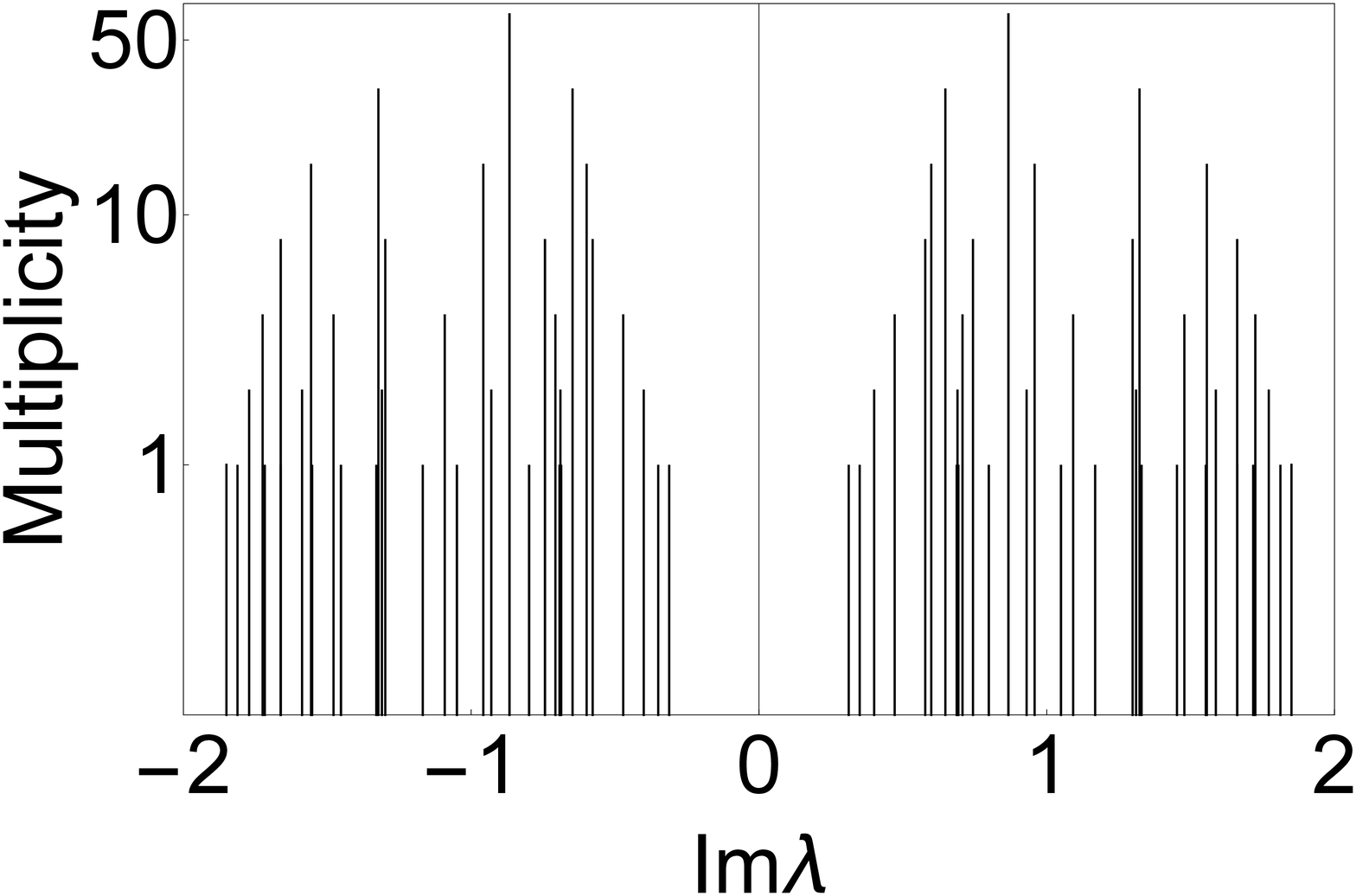}
\caption{ }
\label{fig:eigValDevila}
\end{subfigure}
\begin{subfigure}{8.6cm}
\includegraphics[width=8.6cm]{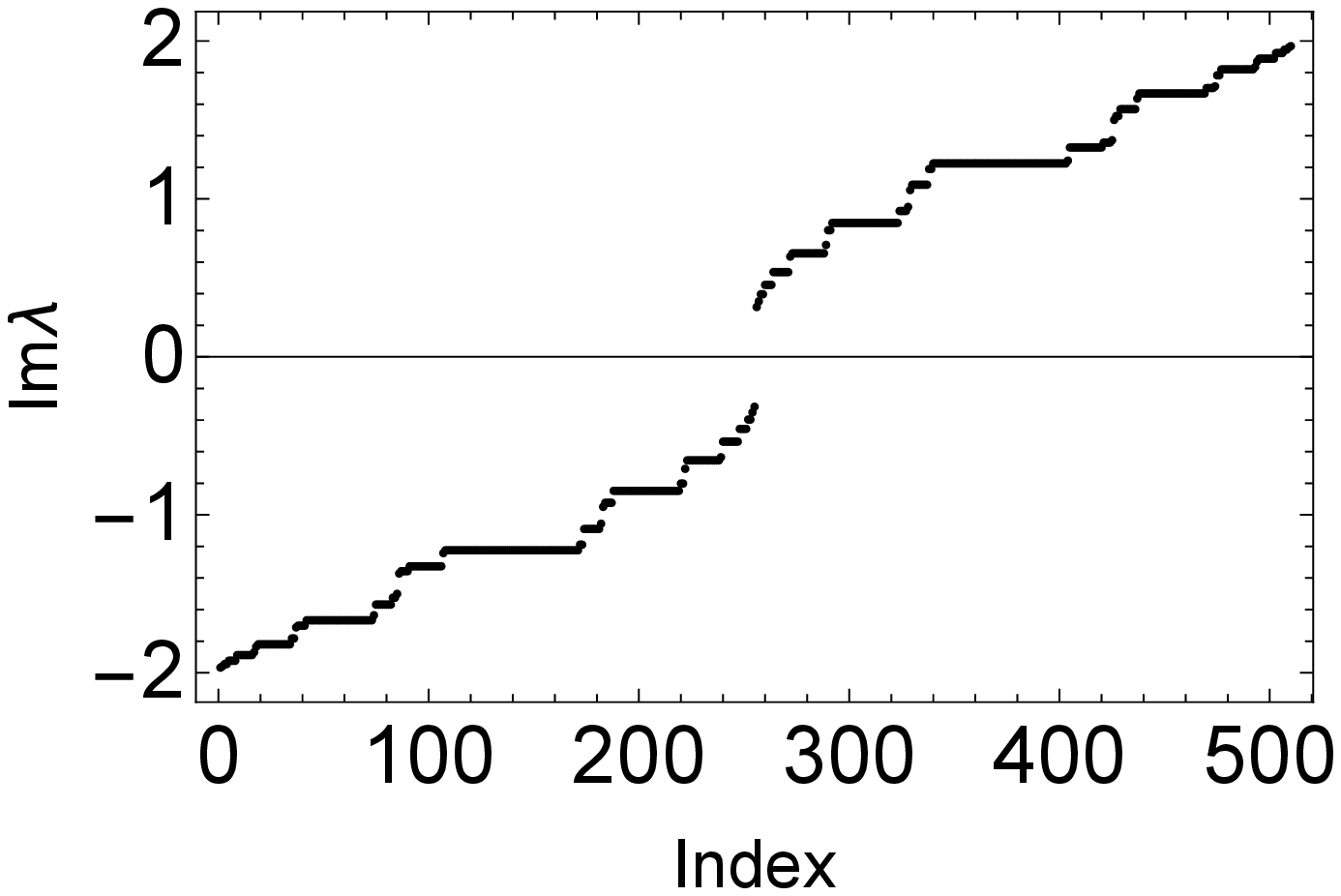}
\caption{ }
\label{fig:eigValDevilb}
\end{subfigure}
\caption{(a) Histogram and (b) eigenvalue
distribution of the conservative 8-level mechanistic model for $\sigma=1/2$.}%
\label{fig:eigValDevil}%
\end{figure}

The eigenvalues of system \eqref{eq:baseMtx} are the roots of the
characteristic equation%
\begin{equation}
P(\lambda)=\det(\mathcal{M}\lambda^{2}+\mathcal{C}\lambda+\mathcal{K}%
)=0.\label{chareqeq}%
\end{equation}
\noindent Figure \ref{fig:eigValDevil} shows the histogram and the eigenvalue
distribution of the purely imaginary eigenvalues of system (\ref{eq:baseMtx})
for the undamped ($\mathcal{C}=0$) case with 8-levels, $\sigma=1/2$.
\noindent The eigenvalue distribution shown in Figure
\ref{fig:eigValDevilb} exhibits a \textquotedblleft devil's
staircase\textquotedblright\ type self-similarity. He et al.
\cite{he2003trees} demonstrated that the adjacency matrix for a class of
symmetric tree graphs have devil's staircase\ type spectrum. Kalm\'{a}r-Nagy
et al. \cite{kalmarnagydevil2017} showed the same for a different type of
self-similar graph, as well as that the characteristic polynomial of the
adjacency matrix of such a graph is a product of Chebyshev polynomials.

\begin{claim}
The characteristic polynomial (\ref{chareqeq}) of the undamped system (Eq.
(\ref{eq:baseMtx}) with $\mathcal{C}=0$) with parameter $\sigma=1/2$ has the
form%
\begin{equation}
P_{0}(\lambda)=1,\quad P_{n}(\lambda)=\frac{Q_{n}(\lambda)}{Q_{n-1}(\lambda
)}P_{n-1}^{2}(\lambda)=Q_{n}(\lambda)\prod_{i=0}^{n-1}Q_{i}(\lambda
)^{2^{n-(i+1)}},\quad n>0.\label{eqrecursive}%
\end{equation}
\noindent Here $Q_{i}(\lambda)$ are generalized Chebyshev-polynomials of the
second kind given by{\small
\begin{equation}
Q_{0}(\lambda)=1,\quad Q_{i}(\lambda)=U_{i}\left(  \frac{\lambda^{2}+2}%
{2}\right)  -\frac{1}{2}U_{i-1}\left(  \frac{\lambda^{2}+2}{2}\right)  ,\quad
i>0,\label{chebyshev}%
\end{equation}
}{\normalsize \noindent}and $U_{i}$ denotes the $i$th Chebyshev-polynomial of
the second kind \cite{mason2002chebyshev}.
\end{claim}

\begin{figure}[h]
\centering
\includegraphics[width=8.6cm]{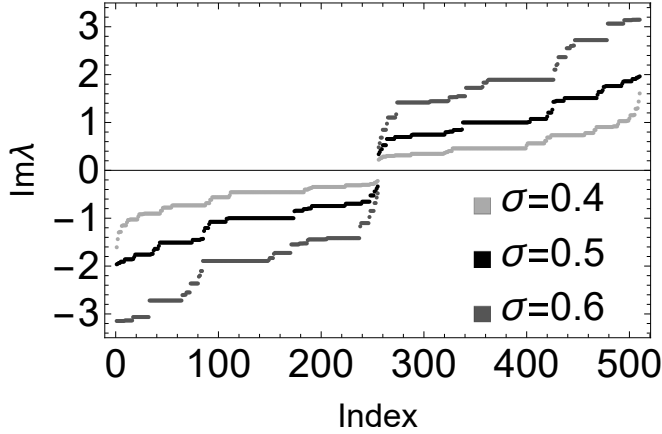} \label{fig:eigValDevil2b}%
\caption{The distribution of $\text{Im}\lambda$ of 8-level damped mechanistic
models for different $\sigma$ values.}%
\label{fig:eigValDevil2}%
\end{figure}

\noindent Figure \ref{fig:eigValDevil2} shows the eigenvalue distributions of
the imaginary parts of the complex eigenvalues for damped 8-level systems.
\noindent These distributions still exhibit similar
characteristics as Figure \ref{fig:eigValDevil}.

\section{The energy spectrum of the mechanistic model}

\label{espectrumsection}

For turbulent flows the energy spectrum $E(\kappa)$ shows the
\textquotedblleft contribution\textquotedblright\ of the different scales of
eddies to the total energy of the flow, i.e. the mean energy stored in the
different wavenumbers $\kappa$. An analogous energy spectrum is now defined
for the mechanistic model showing the energy fraction stored in the different
scales ($M_{l}$) of the system. A consequence of the finite sized mechanistic
turbulence model is that its energy spectrum is discrete, as opposed to the
continuous Kolmogorov spectrum of turbulence.

The total mechanical energy $E(t)$ of the system is the sum of the total
kinetic energy of the masses and the total potential energy stored in the
springs, i.e.
\begin{equation}
E(t)=\frac{1}{2}\dot{\mathbf{x}}^{T}(t)\mathcal{M}\dot{\mathbf{x}}(t)+\frac
{1}{2}\mathbf{x}^{T}(t)\mathcal{K}\mathbf{x}(t).\label{toteq}%
\end{equation}
The total energy is simply the sum of level energies, i.e. $E(t)=\sum
_{l=1}^{n}E_{l}(t)$, where $E_{l}(t)$ is defined so that the potential energy
of the springs connecting two masses are distributed equally between the two
levels and the potential energy of the top spring and the bottom springs are
added to $E_{1}(t)$ and $E_{n}(t)$, respectively:
\begin{equation}%
\begin{split}
E_{l}(t)=\frac{1}{2}M_{l}\sum\limits_{i\in I(l)}\dot{x}_{i}^{2}  &  +\frac
{1}{4}(1+\delta_{l,1})K_{l}\sum\limits_{i\in I(l)}{(x_{i}-x_{\mathcal{P}%
(i)})^{2}}+\\
&  +\frac{1}{4}(1+\delta_{l,n})K_{l+1}\sum\limits_{i\in I(l+1)}{(x_{i}%
-x_{\mathcal{P}(i)})^{2}},\quad l=1,...,n.
\end{split}
\label{levelequations}%
\end{equation}
\noindent Here $I(l)=\{2^{l-1},...,2^{l}-1\}$ are the indices on level $l$ and
$\delta$ is the Kronecker-delta ($\delta_{i,j}=1$ for $i=j$, $0$ otherwise).
This definition of $E_{l}(t)$ ensures its non-negativity for $t\geq0$.

\begin{figure}[h]
\centering
\begin{subfigure}[h]{8.6cm}
\includegraphics[width=8.6cm]{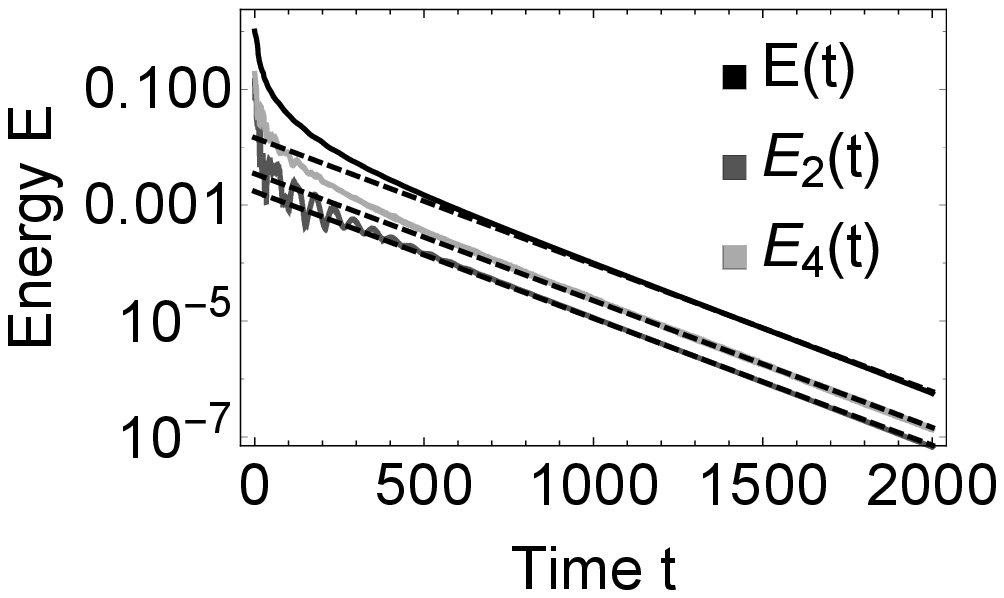}
\caption{}
\label{energiesintimea}
\end{subfigure}
\begin{subfigure}[h]{8.6cm}
\includegraphics[width=8.6cm]{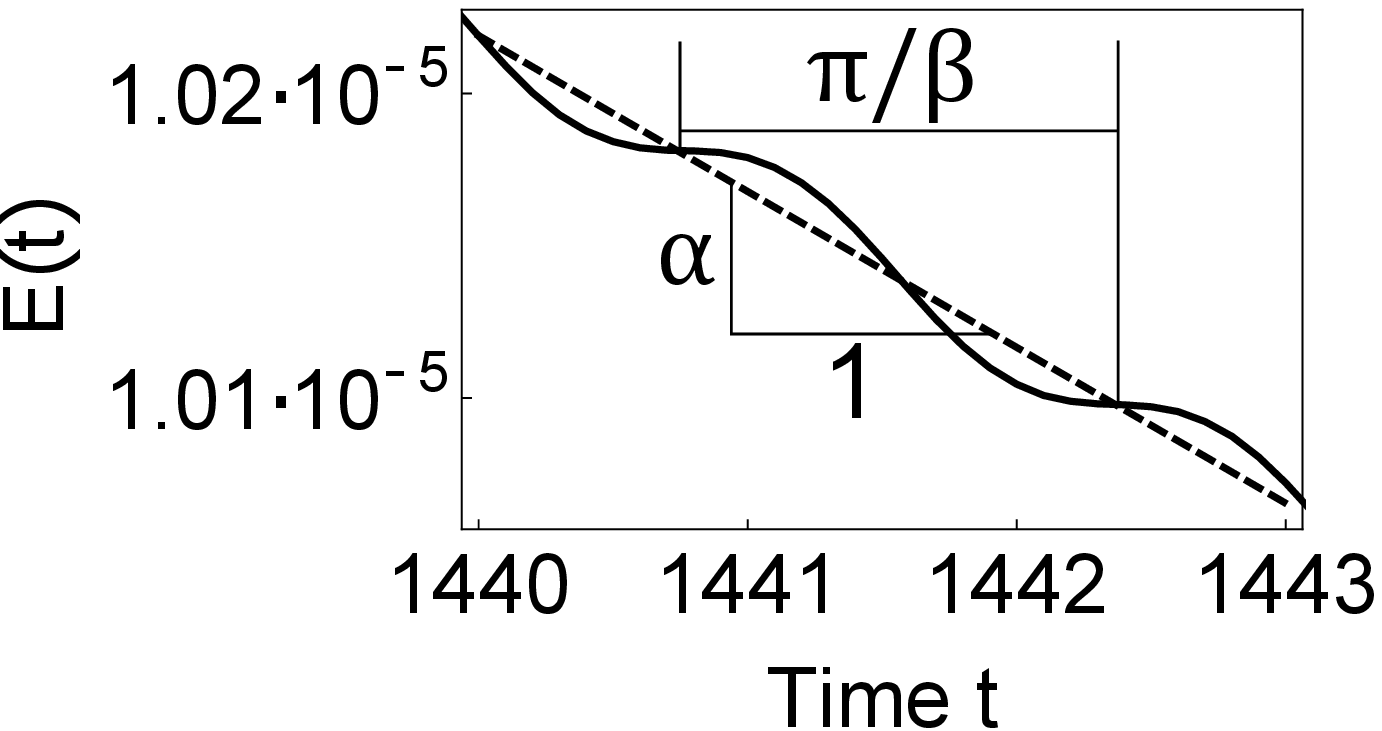}
\caption{}
\label{energiesintimeb}
\end{subfigure}
\caption{(a) The total energy and the energy of the
2nd and 4th levels of an 8-level mechanistic model with $\sigma=1/2$. (b) The
total energy plotted for a narrow time interval.}%
\label{energiesintime}%
\end{figure}

\noindent Figure \ref{energiesintime} shows the total energy and the energy of
the 2nd and 4th levels of an 8-level mechanistic model with $\sigma=1/2$
(launched from random initial conditions). \noindent The behavior of the energies $E(t),E_{l}(t)$
($l=1,...,n$) consists of a transient part and a decaying oscillatory state
(the dashed lines in Figure \ref{energiesintime} are exponentially decaying
functions with the same exponent $2\alpha$). Figure \ref{energiesintimeb} also
shows that the total energy oscillates around the exponentially decaying
function (the dashed line) when $t$ is large. This is the characteristic
behavior of $E(t)$ and each $E_{l}(t)$ ($l=1,...,n$) as $t\rightarrow\infty$.

The asymptotic behavior of the linear system \eqref{eq:baseMtx} and so of its
energy are determined by its rightmost pair of eigenvalues (corresponding to
the slowest decaying motion)
\begin{equation}
\lambda_{slow}=\alpha+\beta i,\quad\lambda_{slow}^{\ast}=\alpha-\beta
i.\ \ \label{dominanteigenvalues}%
\end{equation}
We define the mean level energy $\bar{E}_{l}$ and the mean energy $\bar{E}$ in
the asymptotic limit $\left(  \tau\rightarrow\infty\right)  $ as
\begin{equation}
\bar{E}_{l}=\frac{\beta}{\pi}\int_{\tau}^{\tau+\pi/\beta}E_{l}(t)dt,\quad
\bar{E}=\sum_{l=1}^{n}\bar{E}_{l}=\frac{\beta}{\pi}\int_{\tau}^{\tau+\pi
/\beta}E(t)dt,\quad l=1,...,n.\label{integrate2}%
\end{equation}
Here $\pi/\beta$ is the oscillation period of $E(t),\ E_{l}(t)$.

The energy fraction stored in mass wavenumber $\kappa_{l}$ is defined as
\begin{equation}
\hat{E}_{l}=\bar{E}_{l}/\bar{E},\quad l=1,...,n.\label{espectrumwithintegrals}%
\end{equation}
\noindent This energy fraction expresses the contribution\ of a scale
(signified by the mass wavenumber) to the total energy of the system. The
$\hat{E}_{l}$ values constitute the discrete energy spectrum $\hat{E}$ of the
mechanistic model.

\subsection{Derivation of the energy spectrum}

\label{derivationofespectrum}

Eq. \eqref{eq:baseMtx} can be recast as a system of $2N$ first order
differential equations
\begin{equation}
\dot{\mathbf{y}}(t)=A\mathbf{y}(t),\quad\mathbf{y}(0)=\mathbf{y}%
_{0}.\label{2neq}%
\end{equation}
With the decomposition \cite{nakic2002optimal}
\begin{equation}
D_{1}D_{1}^{\ast}=\mathcal{K},\quad D_{2}D_{2}^{\ast}=\mathcal{M}%
,\label{transformedbaseeq}%
\end{equation}%
\begin{equation}
A=%
\begin{pmatrix}
0 & D_{1}^{\ast}D_{2}^{-\ast}\\
-D_{2}^{-1}D_{1} & -D_{2}^{-1}\mathcal{C}D_{2}^{-\ast}%
\end{pmatrix}
,\quad\mathbf{y}(t)=%
\begin{pmatrix}
D_{1}^{\ast}\mathbf{x}(t)\\
D_{2}^{\ast}\dot{\mathbf{x}}(t)
\end{pmatrix}
,\quad\mathbf{y}_{0}=%
\begin{pmatrix}
D_{1}^{\ast}\mathbf{x}_{0}\\
D_{2}^{\ast}\mathbf{v}_{0}%
\end{pmatrix}
.\label{transformedbaseeq}%
\end{equation}
Here $\ast$, $-\ast$ denotes the conjugate transpose matrix and its inverse,
respectively. \noindent The solution $\mathbf{y}(t)$ of \eqref{2neq} is
\begin{equation}
\mathbf{y}(t)=e^{At}\mathbf{y}_{0}.\label{ysolution}%
\end{equation}
\noindent The total energy of the system is
\begin{equation}
E(t)=\frac{1}{2}||\mathbf{y}(t)||^{2}=\frac{1}{2}\mathbf{y}_{0}^{T}e^{A^{\ast
}t}e^{At}\mathbf{y}_{0}.\label{totewithy}%
\end{equation}
\noindent Since we are interested in the asymptotic behavior of the system,
only the terms related to the rightmost eigenvalue pair $\lambda
_{slow},\ \lambda_{slow}^{\ast}=\alpha\pm\beta i$ of $A$ count. The general
solution of Eq. \eqref{2neq} in the asymptotic limit is
\begin{equation}
\mathbf{y}(t)\cong e^{\alpha t}\big[\mathbf{b}_{1}\text{cos}(\beta
t)+\mathbf{b}_{2}\text{sin}(\beta t)\big],\quad t\rightarrow\infty
\label{dominantmotionsolution2}%
\end{equation}
\noindent where $\mathbf{b}_{1},\mathbf{b}_{2}$ are constant vectors depending
on the initial condition $\mathbf{y}_{0}$. Substituting
\eqref{dominantmotionsolution2} into \eqref{totewithy} leads to
\begin{equation}
E(t)\cong e^{2\alpha t}\left[  B_{1}+B_{2}\text{cos}(2\beta t)+B_{3}%
\text{sin}(2\beta t)\right]  ,\quad t\rightarrow\infty
,\label{simplesolutionforE}%
\end{equation}
\noindent where $B_{1}$, $B_{2}$, $B_{3}$ are initial condition dependent
constants. The decay $2\alpha$ and the frequency $2\beta$ are due to the
quadratic form of \eqref{totewithy}. The energy of each level is described by
a similar expression with different constants $B_{l,1}$, $B_{l,2}$ and
$B_{l,3}$, i.e.
\begin{equation}
E_{l}(t)\cong e^{2\alpha t}\left[  B_{l,1}+B_{l,2}\text{cos}(2\beta
t)+B_{l,3}\text{sin}(2\beta t)\right]  ,\quad l=1,...,n,\quad t\rightarrow
\infty.\label{simplesolutionforEl}%
\end{equation}
\noindent To perform the averaging in Eq. \eqref{integrate2}, we substitute
\eqref{simplesolutionforE} and \eqref{simplesolutionforEl} into
\eqref{integrate2}. The result is
\begin{equation}%
\begin{split}
\bar{E}=\bigg[e^{2\alpha t}\bigg(\frac{B_{1}}{2\alpha}+\frac{\alpha
B_{2}-\beta B_{3}}{2\left(  \alpha^{2}+\beta^{2}\right)  }  &  \text{cos}%
(2\beta t)+\frac{\beta B_{2}+\alpha B_{3}}{2\left(  \alpha^{2}+\beta
^{2}\right)  }\text{sin}(2\beta t)\bigg)\bigg]_{\tau}^{\tau+\frac{\pi}{\beta}%
}=\\
&  =\frac{e^{2\alpha\tau}\left(  e^{2\alpha\frac{\pi}{\beta}}-1\right)  }%
{2}\bigg(\frac{B_{1}}{\alpha}+\frac{\alpha B_{2}-\beta B_{3}}{\left(
\alpha^{2}+\beta^{2}\right)  }\bigg),
\end{split}
\label{simplesolutionforElkappa}%
\end{equation}%
\begin{equation}
\bar{E}_{l}=\frac{e^{2\alpha\tau}\left(  e^{2\alpha\frac{\pi}{\beta}%
}-1\right)  }{2}\bigg(\frac{B_{l,1}}{\alpha}+\frac{\alpha B_{l,2}-\beta
B_{l,3}}{\left(  \alpha^{2}+\beta^{2}\right)  }\bigg),\quad
l=1,...,n.\label{simplesolutionforEbar}%
\end{equation}
\noindent The energy spectrum is obtained by substituting
\eqref{simplesolutionforElkappa} and \eqref{simplesolutionforEbar} into
\eqref{espectrumwithintegrals}
\begin{equation}
\hat{E}_{l}=\frac{(\alpha^{2}+\beta^{2})B_{l,1}+\alpha^{2}B_{l,2}-\alpha\beta
B_{l,3}}{(\alpha^{2}+\beta^{2})B_{1}+\alpha^{2}B_{2}-\alpha\beta B_{3}},\quad
l=1,...,n.\label{finalenergyspectrum}%
\end{equation}
\noindent To calculate the coefficients $B_{1},\ B_{2},\ B_{3}$, and
$B_{l,1},\ B_{l,2},\ B_{l,3}$, we specify the initial condition $\mathbf{y}%
_{0}$ in the two-dimensional "slow" subspace of the $2N$-dimensional state
space of Eq. (\ref{2neq}). For example we can take
\begin{equation}
\mathbf{y}_{0}=\text{Re}\mathbf{s}_{slow}.\label{specialic}%
\end{equation}
\noindent For this initial condition, the evolution described by Eq.
\eqref{simplesolutionforE} is exact, i.e.
\begin{equation}
E(t)=\frac{1}{2}\mathbf{y}_{0}^{T}e^{A^{\ast}t}e^{At}\mathbf{y}_{0}=e^{2\alpha
t}\big[B_{1}+B_{2}\text{cos}(2\beta t)+B_{3}\text{sin}(2\beta t)\big],\quad
t\geq0.\label{toteforspecic}%
\end{equation}
\noindent The unknown constants $B_{1},\ B_{2},\ B_{3}$ can be computed from
the system of algebraic equations
\begin{equation}%
\begin{tabular}
[c]{l}%
$E(0)=\frac{1}{2}\mathbf{y}_{0}^{T}\mathbf{y}_{0}=B_{1}+B_{2}$,\vspace{2mm}\\
$\dot{E}(0)=\frac{1}{2}\mathbf{y}_{0}^{T}(A^{\ast}+A)\mathbf{y}_{0}=2\alpha
B_{1}+2\alpha B_{2}+2\beta B_{3}$,\vspace{2mm}\\
$\ddot{E}(0)=\frac{1}{2}\mathbf{y}_{0}^{T}(A^{\ast2}+2A^{\ast}A+A^{2}%
)\mathbf{y}_{0}=4\alpha^{2}B_{1}+(4\alpha^{2}-4\beta^{2})B_{2}+8\alpha\beta
B_{3}$,
\end{tabular}
\ \ \ \ \ \ \ \label{iceqenergies}%
\end{equation}
\noindent created by taking the first and second derivatives of both the LHS
and the RHS of Eq. \eqref{toteforspecic} at $t=0$. \noindent The solution is
\begin{equation}%
\begin{pmatrix}
B_{1}\\
B_{2}\\
B_{3}%
\end{pmatrix}
=\frac{1}{4\beta^{2}}%
\begin{pmatrix}
4(\alpha^{2}+\beta^{2})E(0)-4\alpha\dot{E}(0)+\ddot{E}(0)\\
-4\alpha^{2}E(0)+4\alpha\dot{E}(0)-\ddot{E}(0)\\
-4\alpha\beta E(0)+2\beta\dot{E}(0)
\end{pmatrix}
.\label{bvalues}%
\end{equation}
\noindent The constants $B_{l,1},\ B_{l,2},\ B_{l,3}$ ($l=1,...,n$) are
calculated similarly:
\begin{equation}%
\begin{pmatrix}
B_{l,1}\\
B_{l,2}\\
B_{l,3}%
\end{pmatrix}
=\frac{1}{4\beta^{2}}%
\begin{pmatrix}
4(\alpha^{2}+\beta^{2})E_{l}(0)-4\alpha\dot{E}_{l}(0)+\ddot{E}_{l}(0)\\
-4\alpha^{2}E_{l}(0)+4\alpha\dot{E}_{l}(0)-\ddot{E}_{l}(0)\\
-4\alpha\beta E_{l}(0)+2\beta\dot{E}_{l}(0)
\end{pmatrix}
.\label{blvalues}%
\end{equation}
\noindent The initial values $E(0),\ \dot{E}(0)$ and $\ddot{E}(0)$ are
calculated from the quadratic forms in Eq. \eqref{iceqenergies}, while
$E_{l}(0)$, $\dot{E}_{l}(0),\ \ddot{E}_{l}(0)$ are calculated from Eq.
\eqref{levelequations} evaluated at $t=0$. The initial conditions
$\mathbf{x}_{0},\mathbf{v}_{0}$ are calculated from $\mathbf{y}_{0}$ using
\begin{equation}%
\begin{pmatrix}
\mathbf{x}_{0}\\
\mathbf{v}_{0}%
\end{pmatrix}
=%
\begin{pmatrix}
D_{1}^{-\ast}\mathbf{y}_{01}\\
D_{2}^{-\ast}\mathbf{y}_{02}%
\end{pmatrix}
,\label{xvic}%
\end{equation}
\noindent where $\mathbf{y}_{01}$ and $\mathbf{y}_{02}$ are the first $N$ and
last $N$ elements of the $2N$-dimensional $\mathbf{y}_{0}$, respectively.
Substituting \eqref{bvalues} and \eqref{blvalues} into
\eqref{finalenergyspectrum} leads to
\begin{equation}
\hat{E}_{l}=\frac{\ddot{E}_{l}(0)-6\alpha\dot{E}_{l}(0)+4E_{l}(0)\left(
3\alpha^{2}+\beta^{2}\right)  }{\ddot{E}(0)-6\alpha\dot{E}(0)+4E(0)\left(
3\alpha^{2}+\beta^{2}\right)  }.\label{espectrumformula}%
\end{equation}
Even though we can now compute the energy spectrum, its calculation is only
feasible for relatively low number ($\sim10$) of levels due to the rapidly
increasing size of matrix $A$ (an additional level almost doubles the number
of masses in the mechanistic model). To overcome this issue, we reduce the
mechanistic model by replacing entire levels of masses with single
representative masses which leads to an equivalent chain oscillator.

\subsection{The energy spectrum for large mechanistic models - The equivalent
chain oscillator}

\label{eqchainosc}

Our approach is similar in spirit to renormalization techniques
\cite{binney1992theory}. Renormalization group theory is extensively used to
decimate the elements of large vibrational systems/chain oscillators
\cite{keirstead1988vibrational,hastings2003random}. We group the masses
located on the same level to represent an entire level with a single mass,
reducing the mechanistic model to a chain oscillator (Figure
\ref{mechmodandchainoscb}). In general, chain oscillators are the simplest
mechanical systems which consist of a chain of masses connected by springs
(and dampers) \cite{wang1973propagation,santos1990spring,kresimir2011damped}.
The equivalent chain oscillator should exhibit the same energy spectrum (the
energy distribution among the masses in the chain) as the mechanistic
turbulence model.

\begin{figure}[h]
\centering
\begin{subfigure}[h]{8.6cm}
\includegraphics[width=8.6cm]{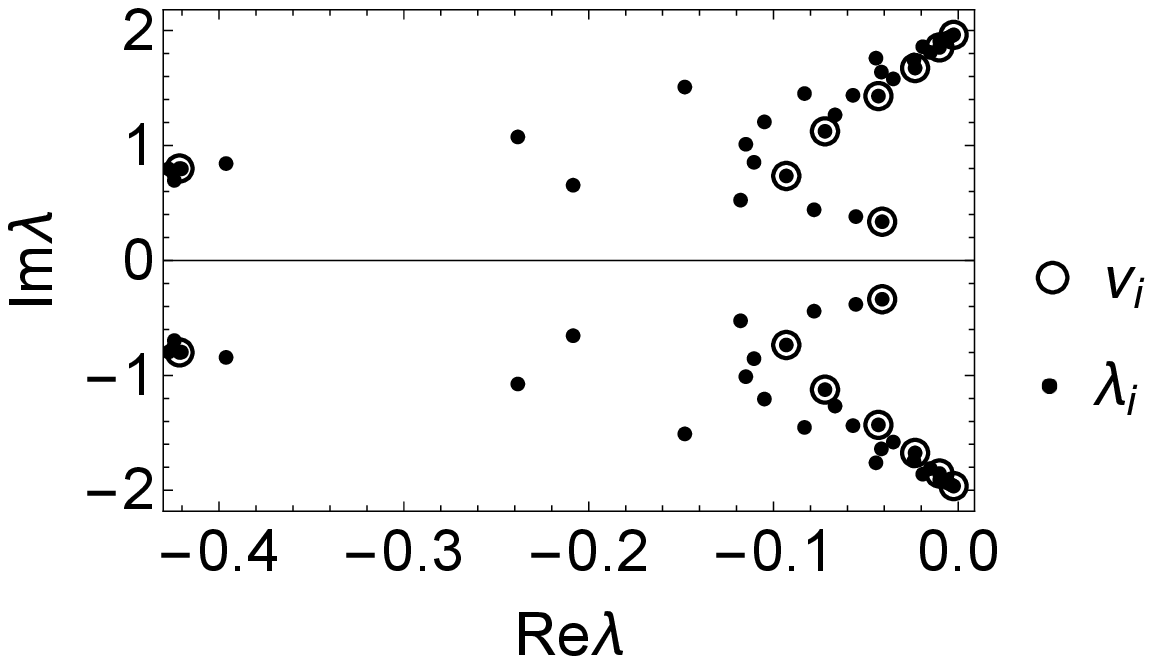}
\caption{}
\label{eigenvaluesbothmodelsa}
\end{subfigure}
\begin{subfigure}[h]{8.6cm}
\includegraphics[width=8.6cm]{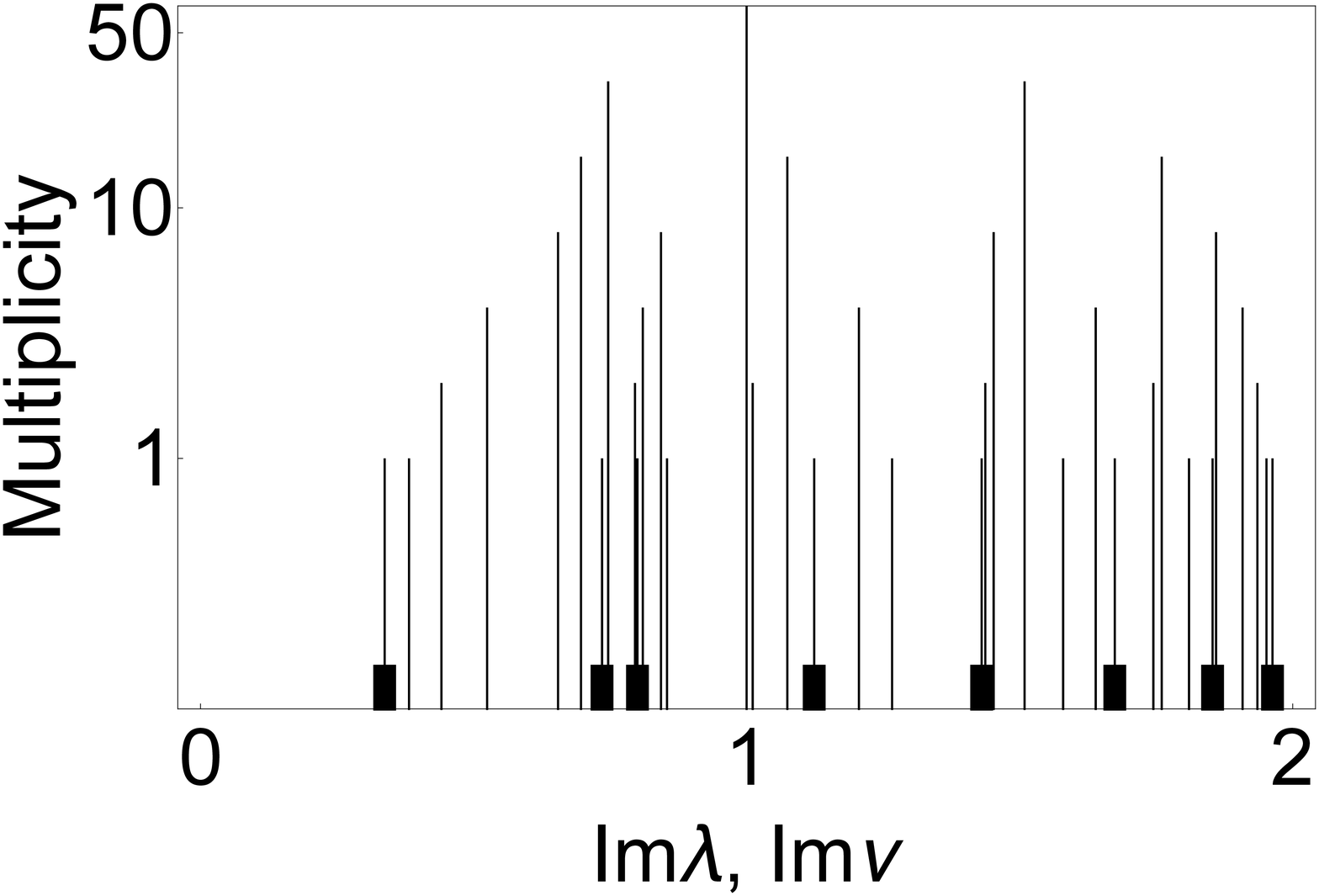}
\caption{}
\label{eigenvaluesbothmodelsb}
\end{subfigure}
\begin{subfigure}[h]{8.6cm}
\includegraphics[width=8.6cm]{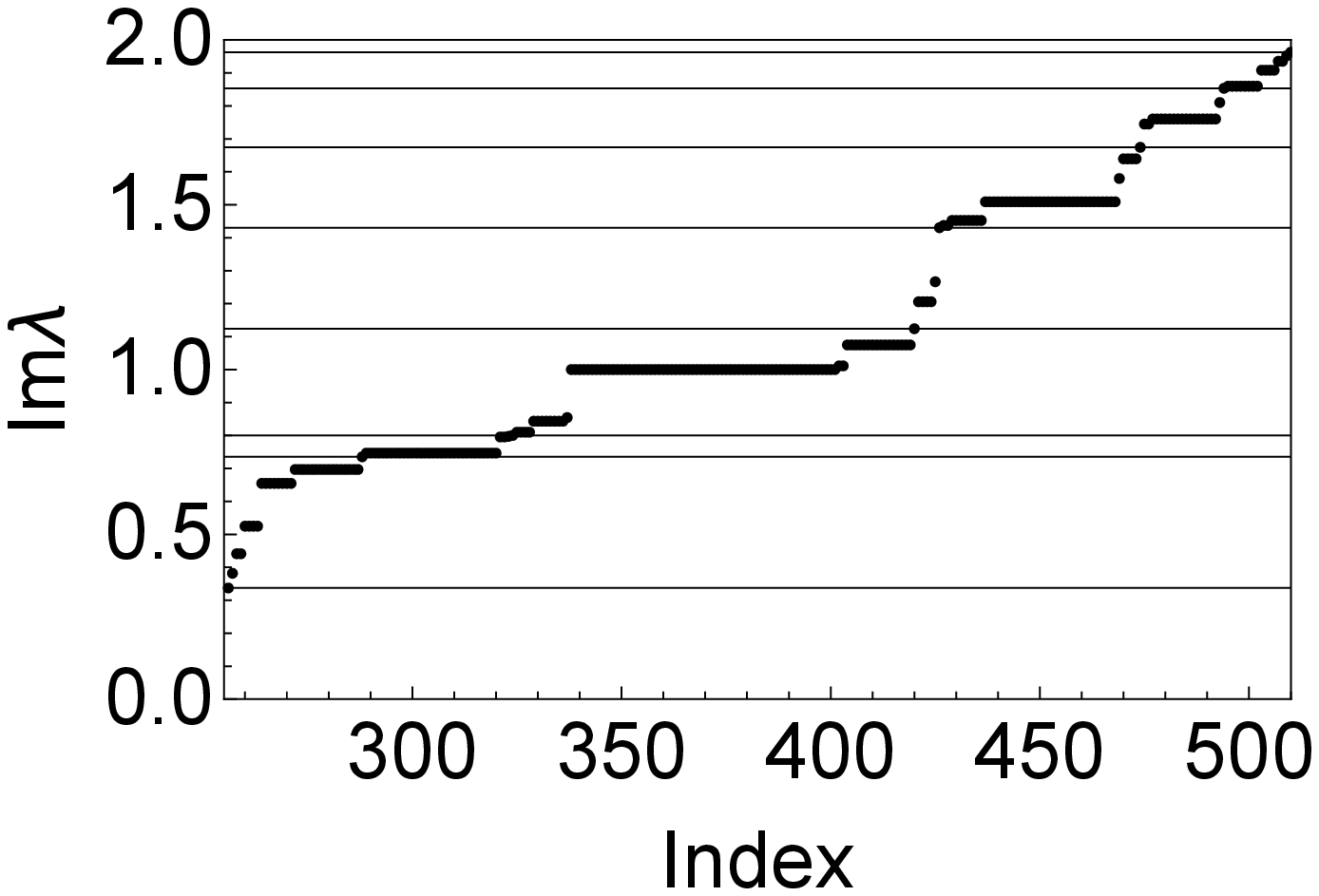}
\caption{}
\label{eigenvaluesbothmodelsc}
\end{subfigure}
\caption{(a) Eigenvalues $\lambda_{i}$ and
$\nu_{i}$ for $\sigma=1/2$. (b) Histogram of $\text{Im}\lambda_{i}$ for
$\sigma=1/2$. The wide columns mark $\text{Im}\nu_{i}$, each has multiplicity
of one. (c) Eigenvalue distribution of $\text{Im}\lambda_{i}$ for $\sigma
=1/2$, the horizontal lines mark $\text{Im}\nu_{i}$.}%
\label{eigenvaluesbothmodels}%
\end{figure}

\noindent The elements of the mechanistic model are connected in parallel
within a level, thus it can be shown (see Eqs. \eqref{powerlawm},
\eqref{powerlawk}, \eqref{dimdamplarge}) that the parameters of the equivalent
chain oscillator are
\begin{equation}%
\begin{tabular}
[c]{c}%
$M_{l}^{\prime}=2^{l-1}M_{l}=1,\quad l=1,...,n$,\vspace{1mm}\\
$K_{l}^{\prime}=2^{l-1}K_{l},\quad l=1,...,n+1$,\vspace{1mm}\\
$C_{n+1}^{\prime}=2^{n}C_{n+1}$,
\end{tabular}
\ \ \ \ \ \ \ \label{chainoscparams}%
\end{equation}
\noindent thus the matrix form of the equations of motion is%
\begin{equation}
\ddot{\mathbf{z}}+\mathcal{C}^{\prime}\dot{\mathbf{z}}+\mathcal{K}^{\prime
}\mathbf{z}=\mathbf{0},\quad\mathbf{z}(0)=\mathbf{z}_{0},\quad\mathbf{\dot{z}%
}(0)=\mathbf{u}_{0}.\label{eq:baseMtxchain}%
\end{equation}
\noindent The damping matrix $\mathcal{C}^{\prime}$ has only one nonzero
element, $\mathcal{C}_{n,n}^{\prime}=C_{n+1}^{\prime}=2^{\frac{n-1}{2}}%
\sigma^{\frac{n}{2}}$ (see Eq. \eqref{dimdamplarge}) and the stiffness matrix
$\mathcal{K}^{\prime}$ is a tridiagonal matrix: $\mathcal{K}_{l,l}^{\prime
}=K_{l}^{\prime}+K_{l+1}^{\prime}=2^{l-1}\sigma^{l-1}(1+2\sigma)$
($l=1,...,n$), $\mathcal{K}_{l+1,l}^{\prime}=\mathcal{K}_{l,l+1}^{\prime
}=-K_{l+1}^{\prime}=-2^{l}\sigma^{l}$ ($l=1,...,n-1$).\noindent\ Figure
\ref{eigenvaluesbothmodels} depicts the spectrum of an 8-level mechanistic
turbulence model and that of its equivalent chain oscillator with $\sigma
=1/2$. Since the eigenvalues are complex, only $\text{Im}\lambda>0$ are shown
in Figures \ref{eigenvaluesbothmodelsb} and \ref{eigenvaluesbothmodelsc}.

\begin{claim}
The characteristic polynomial of the undamped system (Eq.
(\ref{eq:baseMtxchain}) with $\mathcal{C}^{\prime}=0$) with parameter
$\sigma=1/2$ is
\begin{equation}
Q_{n}(\nu)=U_{n}\left(  \frac{\nu^{2}+2}{2}\right)  -\frac{1}{2}U_{n-1}\left(
\frac{\nu^{2}+2}{2}\right)  ,\quad n>0,\label{eqrecursivechain}%
\end{equation}
\noindent where $U_{i}$ denotes the $i$th Chebyshev-polynomial of the second
kind (as in Claim 1).
\end{claim}

\noindent Claim 1 and Claim 2 imply that ALL eigenvalues of the equivalent
chain oscillator are eigenvalues of the corresponding mechanistic turbulence
model for $\sigma=1/2$ and most importantly the rightmost pair of eigenvalues
$\nu_{slow},\lambda_{slow}$ are equal. \noindent Even though we have no formal
proof, spectrum inclusion seems to be true for the general, $\sigma\neq1/2$ case.

\subsection{Results}

\label{results}

\begin{figure}[h]
\centering
\begin{subfigure}[h]{4.3cm}
\includegraphics[width=4.3cm]{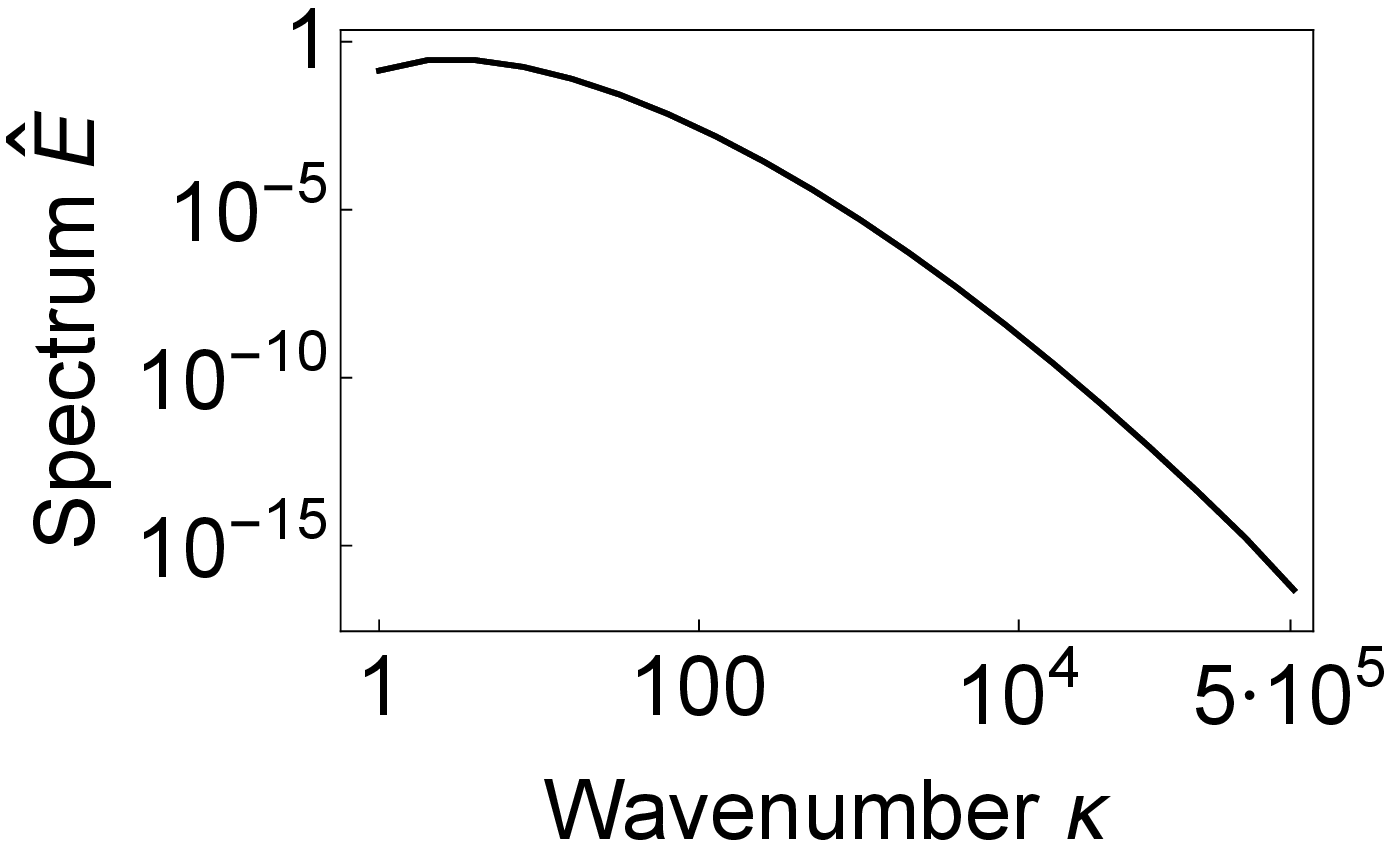}
\caption{$\sigma=0.48$}
\label{espectratimedepa} 
\end{subfigure}
\begin{subfigure}[h]{4.3cm}
\includegraphics[width=4.3cm]{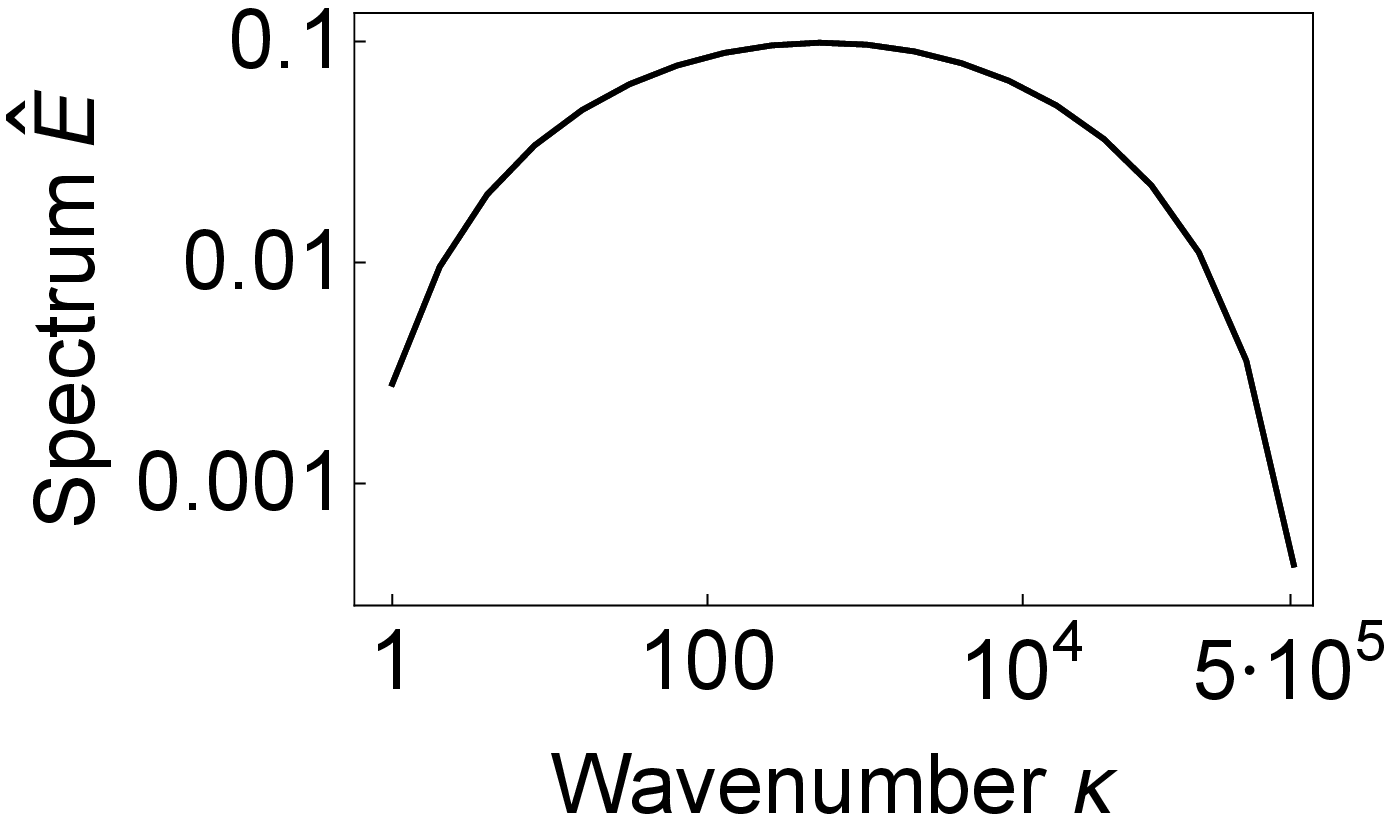}
\caption{$\sigma=0.5$}
\label{espectratimedepb} 
\end{subfigure}
\begin{subfigure}[h]{4.3cm}
\includegraphics[width=4.3cm]{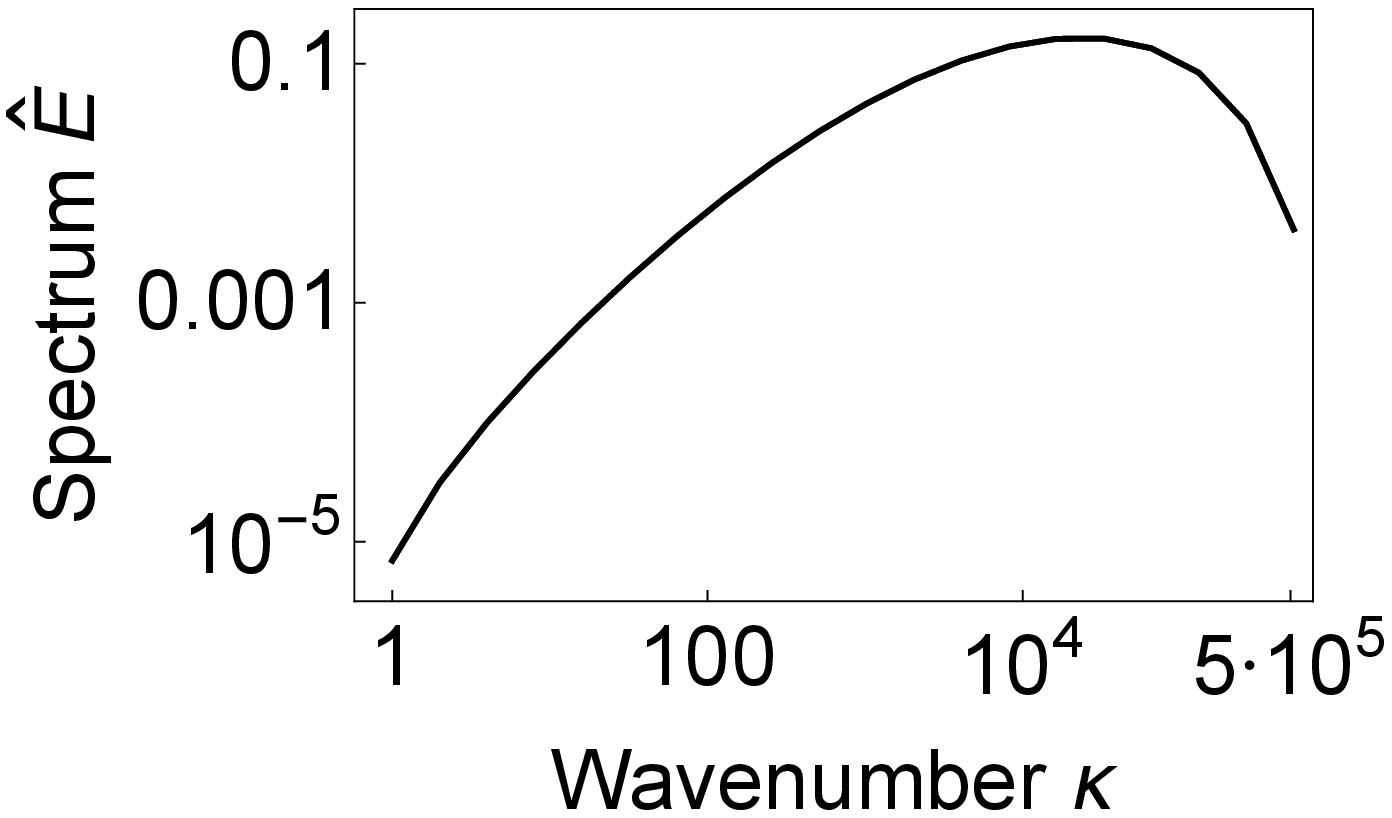}
\caption{$\sigma=0.503$}
\label{espectratimedepc} 
\end{subfigure}
\begin{subfigure}[h]{4.3cm}
\includegraphics[width=4.3cm]{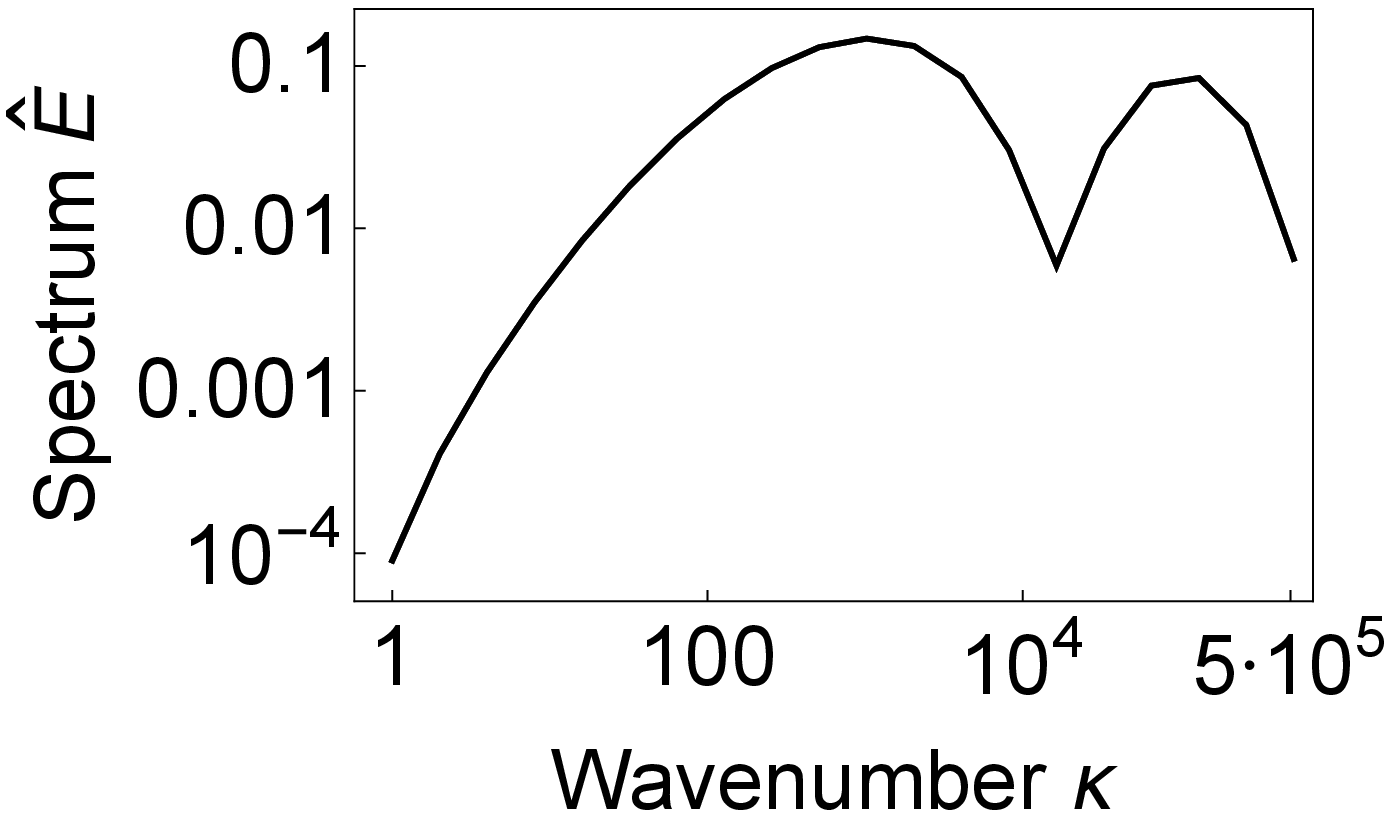}
\caption{$\sigma=0.505$}
\label{espectratimedepd} 
\end{subfigure}
\begin{subfigure}[h]{4.3cm}
\includegraphics[width=4.3cm]{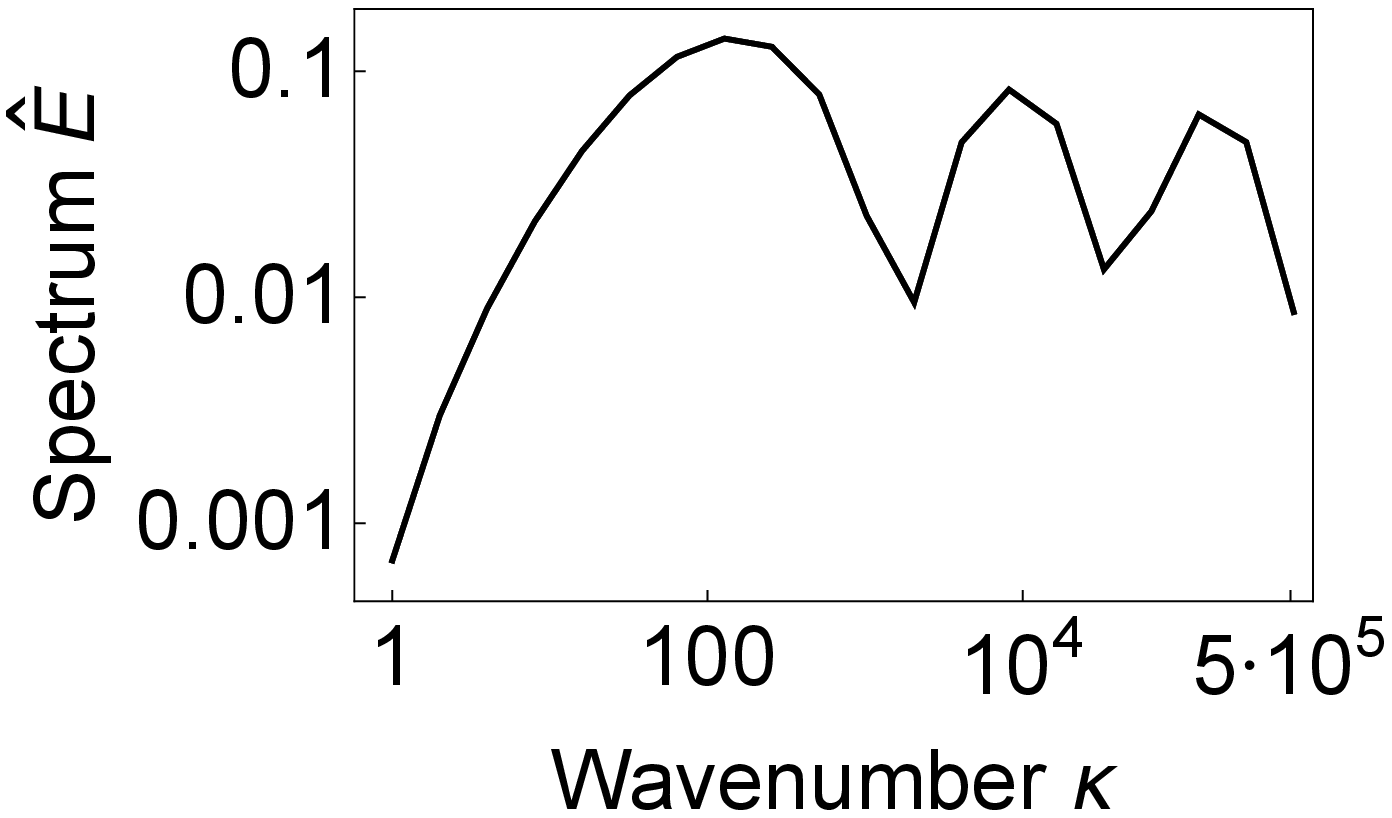}
\caption{$\sigma=0.507$}
\label{espectratimedepe}
\end{subfigure}
\begin{subfigure}[h]{4.3cm}
\includegraphics[width=4.3cm]{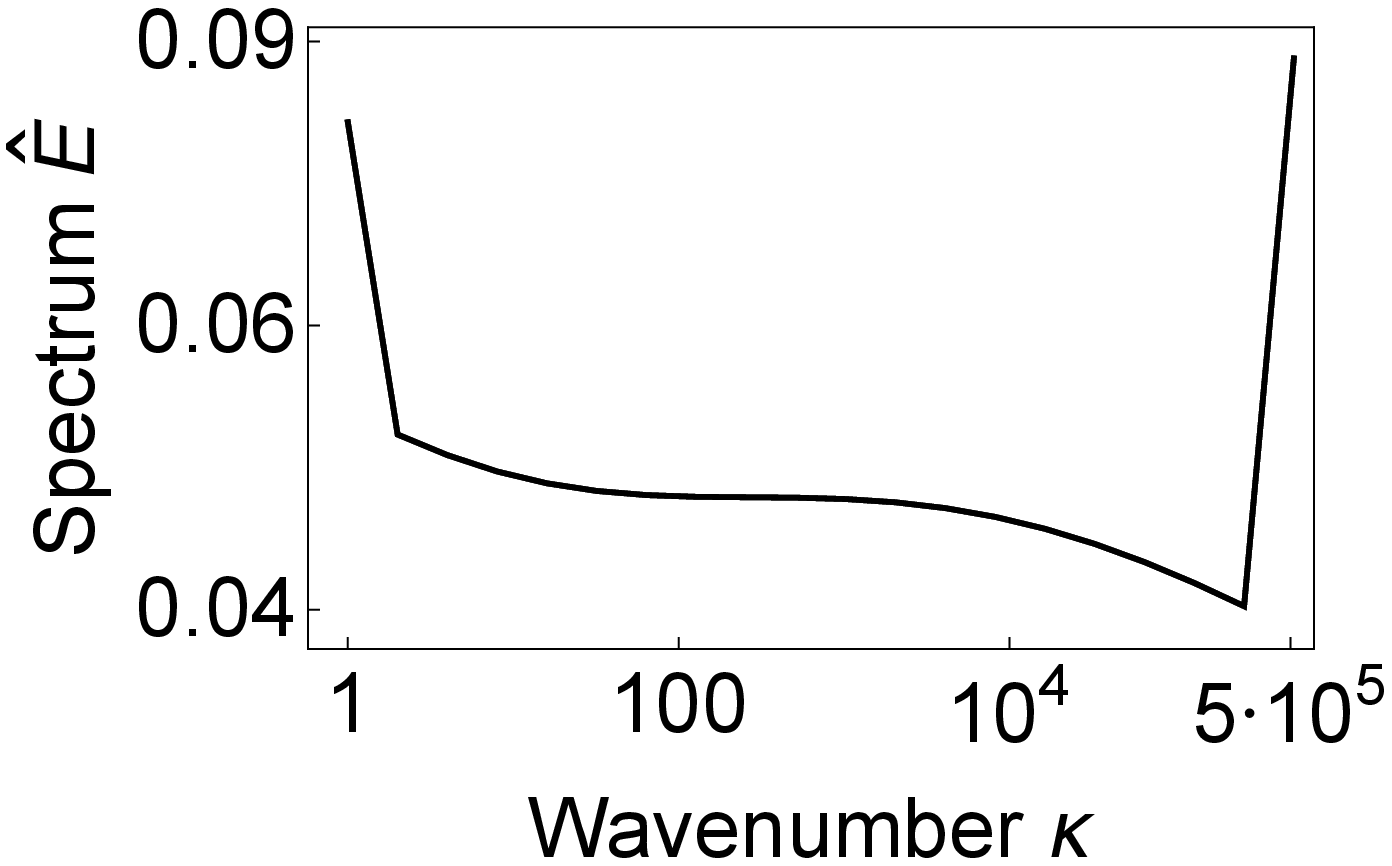}
\caption{$\sigma=0.52$}
\label{espectratimedepf}
\end{subfigure}
\caption{Energy spectra of 20-level mechanistic
models for different $\sigma$'s.}%
\label{espectralinefit}%
\end{figure}

The energy spectrum of the mechanistic model is calculated with different
$\sigma$ values using the equivalent chain oscillator formulation for $n=20$
levels (Figure \ref{espectralinefit}). \noindent For $\sigma<1/2$ (Figure \ref{espectratimedepa},
$\sigma=0.48$) the energy spectrum is similar to the Kolmogorov spectrum (cf.
Figure \ref{kolmogorovspectrum}): the peak of the spectrum is located at a
small wavenumber and the spectrum has a negative slope in the intermediate
range. Figure \ref{espectratimedepb} shows that the energy spectrum is almost
symmetric for $\sigma=1/2$ (the damper attached to the bottom of the chain
breaks the symmetry). Most of the energy is concentrated in the intermediate
scales, the peak is located at the 10th wavenumber. For $\sigma>1/2$ (Figure
\ref{espectratimedepc}, $\sigma=0.503$) the peak of the spectrum moves towards
large wavenumbers and the spectrum has a positive slope in the intermediate
range. The energy spectrum becomes very different as $\sigma$ is further
increased, with local minima appearing (Figures \ref{espectratimedepd} and
\ref{espectratimedepe} with $\sigma=0.505$ and $\sigma=0.507$, respectively).
When $\sigma$ is set even larger (Figure \ref{espectratimedepf}, $\sigma
=0.52$), we get spectra with a significant amount of energy stored in the
smallest and largest wavenumbers.

\begin{figure}[h]
\centering
\begin{subfigure}[h]{0.5\textwidth}
\includegraphics[width=8.6cm]{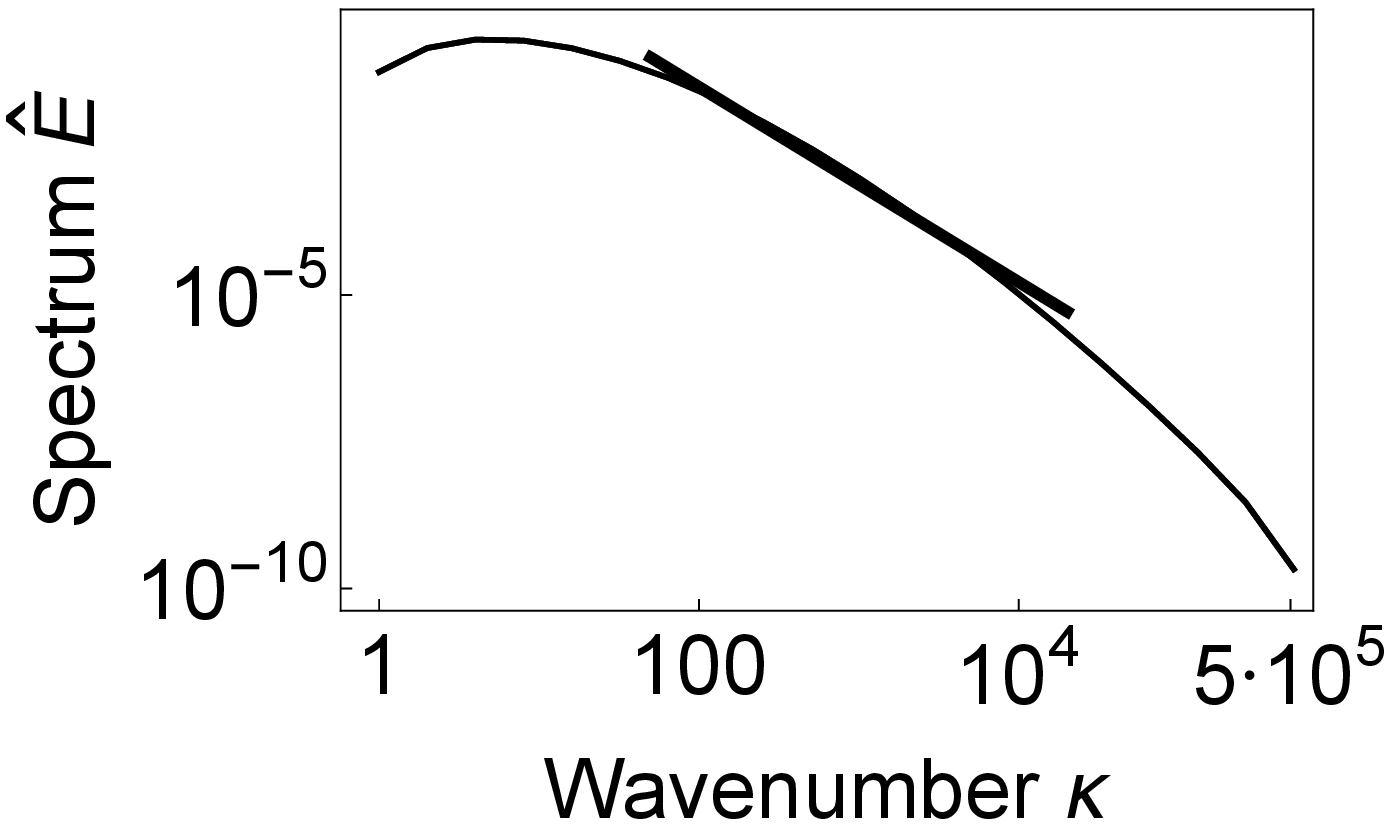}
\caption{}
\label{espectrakolmda} 
\end{subfigure}
\begin{subfigure}[h]{0.5\textwidth}
\includegraphics[width=8.6cm]{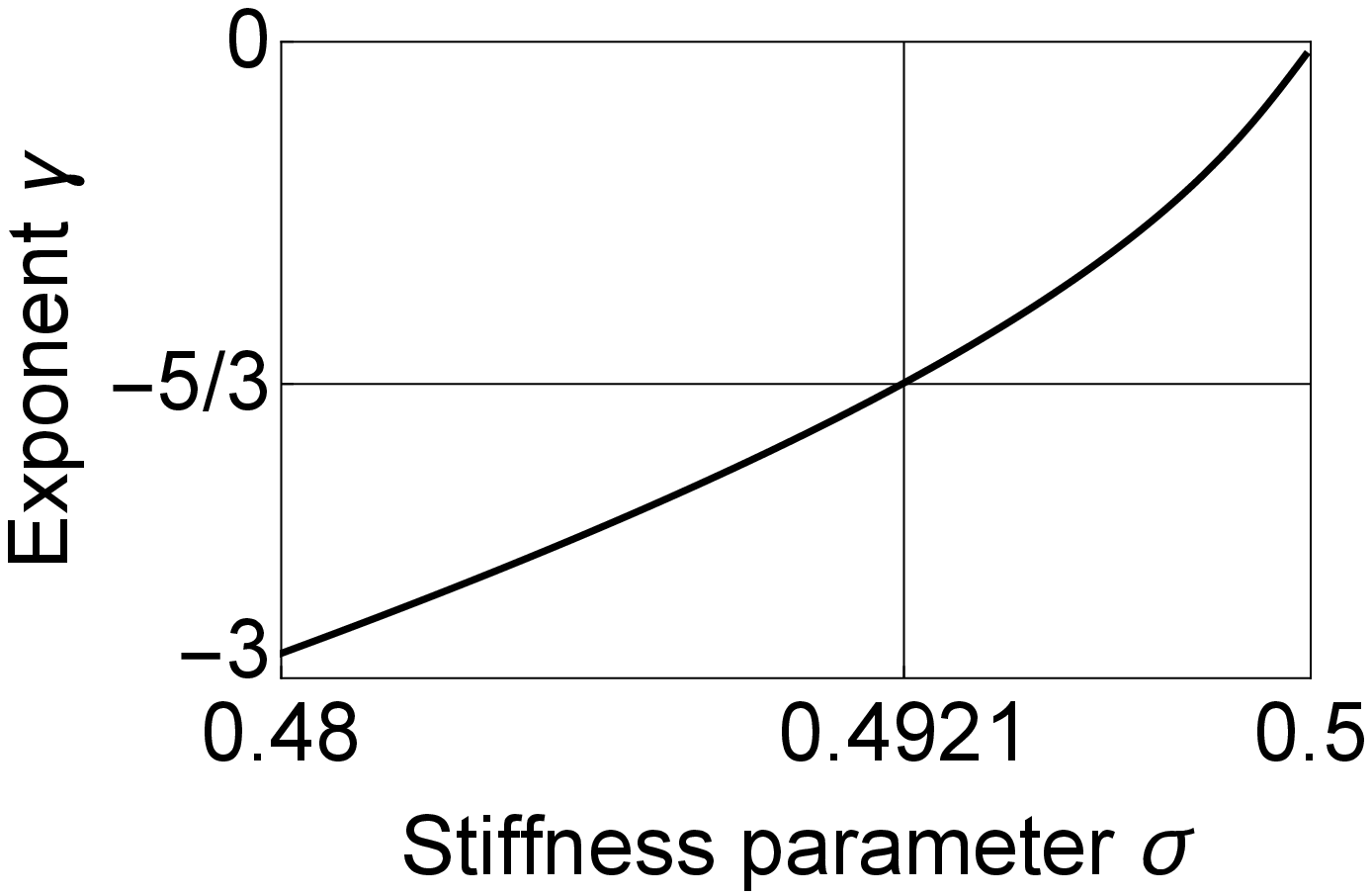}
\caption{}
\label{espectrakolmdb}
\end{subfigure}
\caption{(a) Energy spectrum of the 20-level
mechanistic model for $\sigma=0.4921$. (b) The exponent $\gamma$ as the
function of $\sigma$ for 20-level mechanistic models.}%
\label{espectrakolmd}%
\end{figure}

In the $\sigma<1/2$ case (Figure \ref{espectratimedepa}) we approximate the
intermediate range of the spectrum with the scaling law
\begin{equation}
\hat{E}\sim\kappa^{\gamma}.\label{mechspectrum}%
\end{equation}
\noindent In Figure \ref{espectrakolmdb} the scaling exponent $\gamma$ is
plotted as the function of the stiffness parameter $\sigma$. The main finding
is that $\sigma$ parameter can be tuned to match the Kolmogorov exponent with
$\gamma$: at $\sigma\approx0.4921$ the exponent is $\gamma=-5/3$.
The changes in the shape of the energy spectrum for the
$\sigma>1/2$ cases (Figures \ref{espectratimedepc}, \ref{espectratimedepd},
\ref{espectratimedepe}, and \ref{espectratimedepf}) is due to the change in
the order of the rightmost eigenvalues and the accompanying significant change
of the eigenvector $\mathbf{s}_{slow}$.

\section{Conclusions}

\label{conclusion}

Inspired by Richardson's eddy hypothesis we introduced a mechanistic model of
turbulence with different mass scales. The analysis of the eigenvalues was
performed for different stiffness distributions. The discrete energy spectrum
of the model was defined as the fraction of the total energy stored in the
different mass scales. A formula (Eq. \eqref{espectrumformula}) was derived to
calculate this spectrum in the asymptotic limit. A decimation procedure was
applied to yield an equivalent chain oscillator with energy spectrum identical
to that of the mechanistic chain oscillator. This enabled the calculation of
the energy spectrum for many-level systems. Typical energy spectra were
obtained. In the $\sigma<1/2$ case the energy spectrum is qualitatively
similar to the Kolmogorov spectrum of 3D turbulence. Moreover, we also found
that the stiffness distribution can be tuned so that the exponent $\gamma$ of
the energy spectrum precisely matches with that of the Kolmogorov spectrum.

This work is expected to motivate new studies connecting eigenvalue
distributions of infinite-dimensional linear systems with energy transfer and
resonant capture in nonlinear systems.

\section*{Acknowledgement}

This project was supported by the \'{U}NKP-17-3-I New National Excellence
Program of the Ministry of Human Capacities of Hungary. We acknowledge the
financial support from TeMA Talent Management Foundation.

\bibliographystyle{unsrt}

\end{document}